\documentclass[preprint,12pt]{elsarticle}

 \usepackage{graphicx}
 \usepackage{caption}
 \usepackage{subcaption}
 \usepackage[english]{babel}

\usepackage{amssymb}

\journal{Nuclear Physics A}

\begin{document}

\begin{frontmatter}



\title{X-ray scattering of periodic and graded multilayers: comparison of experiments to simulations from surface microroughness characterization}


\author[addr1]{B.Salmaso}
\fntext[indir]{Phone : +39-039-5971024, Fax: +39-039-5971001}
\ead{bianca.salmaso@brera.inaf.it}

\author[addr1]{D. Spiga}
\author[addr1]{R. Canestrari}
\author[addr2]{L. Raimondi}

\address[addr1]{INAF/Osservatorio Astronomico di Brera, Via E. Bianchi 46, I-23807, Merate - Italy}
\address[addr2]{Sincrotrone Trieste ScpA, S.S. 14 km 163.5 in Area Science Park, 34149 Trieste - Italy}
	  
\begin{abstract}
To enhance the reflectivity of X-ray mirrors beyond the critical angle, multilayer coatings are required. Interface imperfections in the multilayer growth process are known to cause non-specular scattering and degrade the mirror optical performance; therefore, it is important to predict the amount of X-ray scattering from the rough topography of the outer surface of the coating, which can be directly measured, e.g., with an Atomic Force Microscope (AFM). This kind of characterization, combined with X-ray reflectivity measurements to assess the deep multilayer stack structure, can be used to model the layer roughening during the growth process via a well-known roughness evolution model. In this work, X-ray scattering measurements are performed and compared with simulations obtained from the modeled interfacial Power Spectral Densities (PSDs) and the modeled Crossed Spectral Densities for all the couples of interfaces. We already used this approach in a previous work for periodic multilayers; we now show how this method can be extended to graded multilayers. The upgraded code is validated for both periodic and graded multilayers, with a good accord between experimental data and model findings. Doing this, different kind of defects observed in AFM scans are included in the PSD analysis. The subsequent data-model comparison enables us to recognize them as surface contamination or interfacial defects that contribute to the X-ray scattering of the multilayer.
\end{abstract}


\end{frontmatter}


\section{Introduction}
\label{Introduction}
\begin{figure}[!tpb]
        \centering
        \includegraphics[width=0.75\textwidth]{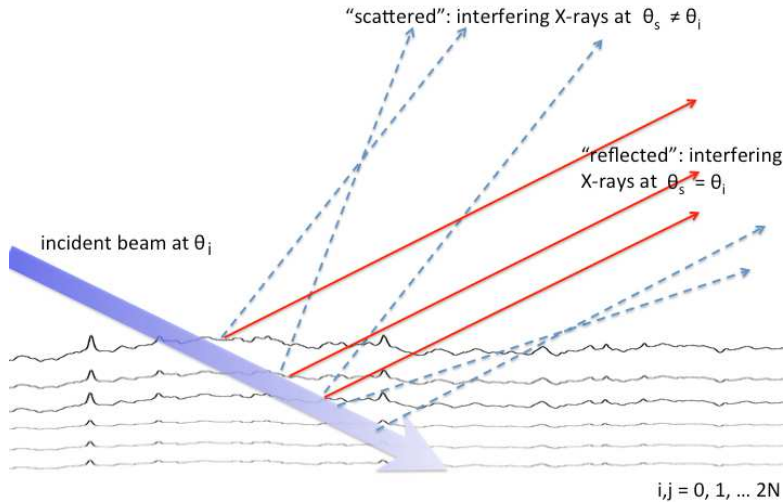}
        \caption{Scheme of X-ray scattering in a multilayer stack. Both "reflected" (i.e., in the direction specular to that of incidence) angle and "scattered" (i.e. in non-specular directions) rays result from the interference of elementary waves scattered at each boundary in the multilayer.}
        \label{fig:schemaXRS}
\end{figure}

Multilayer coatings are known to enhance the reflectivity of extreme ultraviolet (EUV), neutron, and X-ray mirrors at incidence angles larger than the critical one. As the energy of the incident beam increases, the smoothness of the surface becomes more and more important, because the roughness reduces the specular reflectivity and increases the X-ray scattering (XRS) in non-specular directions, leading to a degradation of the angular resolution. Depending on the specific application, periodic or graded multilayers are deposited using different techniques: anyway, to a variable extent, the deposition process triggers an evolution of the roughness from the substrate to the outermost layer. The interference of waves scattered at layer interfaces \cite{Holy 1999} result in the final XRS pattern (Fig.~\ref{fig:schemaXRS}); therefore, in order to estimate the roughness impact on the Point Spread Function (PSF), a roughness measurement of all the multilayer interfaces would be needed.

However, only the outer surface of the multilayer is accessible to direct topography measurements using, e.g., an Atomic Force Microscope (AFM). X-ray reflectivity (XRR) measurements as a function of the incidence angle, combined with a detailed fit routine to interpret the reflectivity scans \cite{Spiga 2007, Spiga 2008}, allows a non-destructive, in-depth analysis of the multilayer stack structure (layer thickness in the stack, uniformity, smoothness), but does not enable the reconstruction of the PSD ({\it Power Spectral Density}) evolving throughout the stack. Nevertheless, an XRS computation based upon the sole thickness description and the outer surface PSD, assuming the rough topography to be exactly replicated in the stack, would in general return a diagram mismatching experimental data (Fig.~\ref{fig:periodico_nogrowth_fields}).

In this paper, a modeling of the PSD evolution in the stack is used to compute the XRS diagram. A known multilayer growth model \cite{Stearns 1998} provides the PSD growth across the stack, modeled from the measured PSD of the substrate and of the multilayer surface. The model physically describes the roughness of each interface as stemming from two effects in mutual competition: the replication of the roughness of the underlying interface and the roughness introduced by the growth of the layer itself (Eq.~\ref{eq:stearns1}). As a result, the PSD increase from the substrate to the outer surface can be modeled by tuning the values of a few growth parameters \cite{Canestrari 2006} that can be tuned to fit the measured external PSD. Once the best-fit parameter values are set, the internal PSDs for all the interfaces and the {\it Crossed Spectral Densities} (CSD) for all the couples of interfaces can be reconstructed. These  quantities, in turn, determine the XRS expected from the multilayer. In particular, it is the cross-correlation between nearby or distant interfaces to affect the coherence between scattered waves, and, consequently, the amplitude of interferential features in the XRS diagram. The first-order perturbation theory (see \cite{Stover 1995, Church 1986}) is used to compute the XRS diagram from the roughness PSD evolution in the stack \cite{Kozhevnikov 2003, Spiga 2005}.  

Previous works \cite{Canestrari 2006, Schroder 2007} have already implemented this model for periodic multilayers. However, broadband multilayers, like the ones in use in X-ray telescopes, have a graded structure~\cite{Tawara 1998}. In this work, we extend the formalism to graded multilayers and we apply it to two multilayer samples, a W/Si periodic sample and a Pt/C graded sample. Roughness analysis of the substrate and the outermost surface of the samples is performed with the Atomic Force Microscope (AFM) operated at INAF/OAB. The layer thickness measurement is obtained from the accurate fit \cite{Spiga 2007} of the XRR measurements performed with a BEDE-D1 diffractometer, also operated at INAF/OAB, at the X-ray energy of 8.045 keV (the Cu K$\alpha_1$ fluorescence line). Eventually, aiming at checking the correctness of the growth description, we have compared the expected XRS diagram to the one measured at with the BEDE-D1 at selected incidence angles, finding a very good agreement between the modeling and the experiment. Some preliminary results were already exposed in a previous paper~\cite{Salmaso 2011}.

In Sect.~\ref{sec:theory} we retrieve the adopted growth model \cite{Stearns 1998} and the XRS formalism applied to multilayers \cite{Spiga 2005}. In Sect.~\ref{sec:experimental} we describe the samples and the experimental setup. In Sect.~\ref{sec:results} we show the modeling of the PSDs growth, as well as the predicted XRS vs. the experiments. For the periodic case, we have reanalyzed the data already treated \cite{Canestrari 2006}, showing that including the modeled PSD growth and adopting a more general electric field modeling leads to the best data/model matching. For the graded case, we show that the thickness trend that describes the stack and fits the XRS peak positions is univocally determined, and also matches the XRR measurement. Finally, we show how XRS can be a powerful tool to discriminate between surface and embedded defects, by including them in the PSD and checking if a proper XRS fit is obtained. The results are briefly summarized in Sect.~\ref{sec:summary}. A possible derivation of the formula used to model the scattering diagram is sketched in \ref{appXRS}, or with more details in \cite{Spiga 2005}. 

\section{Modeling microroughness growth and X-ray scattering in multilayers}
\label{sec:theory}
\subsection{Microroughness growth model}\label{sec:MPES}
The roughness growth model \cite{Stearns 1998} solves a kinetic equation to describe the evolution of the rough profile $z$($x$) with the thickness $\tau$ of the film. For a single layer deposited onto a substrate, this equation reads
\begin{equation}
	\frac{\partial z(x)}{\partial\tau} = -\nu \left|\nabla^n z(x) \right|+\frac{\partial \eta}{\partial \tau}.
	\label{eq:stearns1}
\end{equation}
The model describes the roughening of the surface as a competition between a surface relaxation process and the increase in roughness due to the random nature of the deposition process. The relaxation process is parametrized with $\nu$ and the positive integer $n$ that varies with the kinetic mechanism that dominates the smoothing process \cite{Stearns 1998}. The increase in roughness results from the deposition process and is described by a random shot noise term $\eta$. The solution of Eq.~\ref{eq:stearns1} in terms of surface PSD \cite{Stearns 1998} is
\begin{equation}
	P^{\mathrm{int}}(f)=\Omega \frac {1- \exp(-2\nu {\left| 2\pi f \right|}^n \tau )}{2\nu {\left| 2\pi f \right|}^n},
	\label{eq:Pint}
\end{equation}
where $P^{\mathrm {int}}(f)$ is the intrinsic bi-dimensional PSD of the layer surface, i.e., the PSD that the surface layer would have if the substrate were ideally smooth. This PSD is characterized by a plateau up to the maximum frequency corresponding to the cutoff wavelength $l^* = (\nu\tau)^{\frac{1}{n}}$, then decreases as a power-law of spectral index $n$. $\Omega$ represents the volume of the deposited atom, molecule, or nanocrystal.
         
\begin{figure}[!tpb]
        	\centering
         \includegraphics[width=0.5\textwidth]{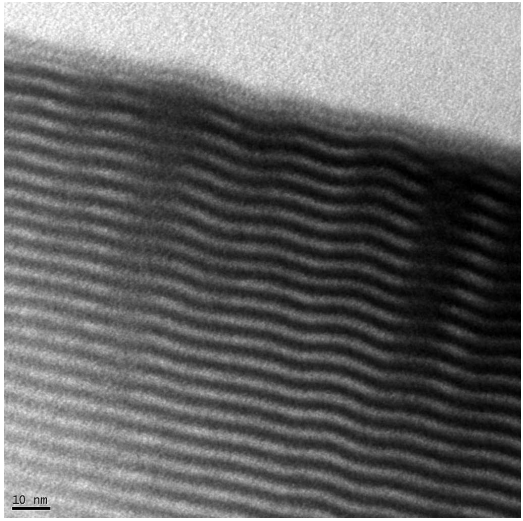}
         \caption{Transmission Electron Microscope section of a W/Si periodic multilayer. The roughness profile is amplified throughout the stack ({\it Courtesy of C. Ferrari and L. Lazzarini, IMEM-CNR}).}
         \label{fig:rough_gr}
\end{figure}

When a stack of $N$ alternated layers is considered, the situation is complicated by the presence of two elements with different properties, i.e., different values of the parameters $\Omega$, $\nu$, and $n$. However, the formalism can be extended by considering each single layer (whose upper surface is labelled with $j$ = 0, 1,\ldots $N$ moving from the substrate towards the surface) as growing upon its underlying layer, which acts as its ÒsubstrateÓ. In this way, one can write \cite{Stearns 1998} $P_0 = P_{\mathrm subs}$ and the PSD of the $j^{\mathrm {th}}$ interface as a sum of the intrinsic contribution of the layer itself and of a term representing the rough profile partially inherited from the previous layer:
\begin{equation}
	P_j(f) = P_j^{\mathrm {int}}(f)+P_j^{\mathrm {ext}}(f) = P_j^{\mathrm{int}}(f)+a_j(f)P_{j-1}(f).
	\label{eq:Pmultilayer}
\end{equation}
In the second term of Eq.~\ref{eq:Pmultilayer}, 
\begin{equation}
	a_j(f) = \exp(-\nu\left|2\pi f\right|^{n}\tau_j)
	\label{eq:replication}
\end{equation}
is a {\it replication factor} that describes to which extent the microrelief components of the $(j-1)^{\mathrm {th}}$ interface are replicated in the $j^{\mathrm {th}}$ layer. Like the $P^{\mathrm {int}}$ term, the replication factor exhibits a cutoff at the spatial wavelength $l^{*} = (\nu\tau)^{\frac{1}{n}}$, that represents the transition above which surface features are damped out. At wavelengths larger than $l^{*}$, the surface topography is almost entirely replicated: the superimposition of the intrinsic term thereby triggers a progressive roughening of the multilayer interface, as a function of $f$ (Fig.~\ref{fig:rough_gr}), to an extent depending on the values of the growth parameters. By fitting these values to the measured PSD growth from the substrate to the outer surface, the parameter values, and consequently the internal PSD evolution, can be determined.

We have translated this formalism into an IDL-based program, firstly aiming at determining the $\Omega_{\mathrm h}$, $\nu_{\mathrm h}$, $n_{\mathrm h}$, $\Omega_{\mathrm l}$, $\nu_{\mathrm l}$, and $n_{\mathrm l}$ for the high- and low-Z density element, respectively. The values that best fit the growth from the substrate to the outer PSD, computed by recursive application of Eq.~\ref{eq:Pmultilayer}, are selected. Once the fitting parameters are found, we compute each $P_j$ by iterating Eq.~\ref{eq:Pmultilayer} $j$ times from the substrate. The CSD between the generic $j^{\mathrm {th}}$ and the $m^{\mathrm {th}}$ interface, with $j < m$ can be simply computed from the replication factors (Eq.~\ref{eq:replication}): 

\begin{equation}
	C_{jm}(f) = a_m (f) \cdot a_{m-1}(f)\cdot \ldots \cdot a_{j+1}(f) P_j (f)
	\label{eq:crosscorr}
\end{equation}

The $P_j$'s and the $C_{jm}$'s are the physical quantities that affect the intensity and the mutual coherence of the scattered waves, which in turn interfere to build up the XRS diagram.

\subsection{X-ray scattering}\label{sec:X-model}
To compute the XRS diagram at the X-ray wavelength $\lambda$ we apply the first order perturbation theory in grazing incidence, assuming the surface to be smooth and isotropic. For a single boundary characterized by a single PSD, the relation is a simple proportionality between the scattered intensity per angle unit and the monodimensional PSD $P(f)$ \cite{Stover 1995, Church 1986},
\begin{equation}
	\frac {1}{I_0} \frac{\mbox{d} I_{\mathrm s} }{\mbox{d} \theta_{\mathrm s }} = \frac{16\pi^2}{\lambda^3} Q_{\mathrm{is}}\,\sin^2\theta_{\mathrm s} \sin\theta_{\mathrm i}\,P(f),
	\label{eq:singleXRS}
\end{equation}
where the scattering angle $\theta_{\mathrm s}$ and the incidence angle $\theta_{\mathrm i}$, measured from the surface (Fig.~\ref{fig:schemaXRS}), are related to the X-ray wavelength $\lambda$ and to the spatial frequency $f$ as follows:
\begin{equation}
	\frac {1}{f} =\frac{\lambda}{|\cos\theta_{\mathrm i}-\cos\theta_{\mathrm s}|}.
	\label{eq:grating}
\end{equation}
The $Q_{\mathrm{is}}$ {\it polarization factor} can be related to the reflectivity of the boundary $R_{\mathrm F}$, computed via the Fresnel equations,
\begin{equation}
	Q_{\mathrm{is}}= \sqrt{R_{\mathrm F}(\theta_{\mathrm s}) \,R_{\mathrm F}(\theta_{\mathrm i})}.
	\label{eq:polarization}
\end{equation}

In the case of a stack of alternated $N$ layers, this relation is complicated by the interference effects among interfaces \cite{Holy 1999, Kozhevnikov 2003}. If the roughness is isotropic and the incidence angle is beyond the critical angles for total reflection of both materials, the scattering diagram can be expressed in terms of the monodimensional $P_j$'s and $C_{jm}$'s using the following equation \cite{Spiga 2005}, which represents a generalization of Eq.~\ref{eq:singleXRS} (see \ref{appXRS}):
\begin{equation}
\frac {1}{I_0} \frac{\mbox{d} I_{\mathrm s}}{\mbox{d}\theta_{\mathrm s}} = K(\lambda, \theta_{\mathrm s}, \theta_{\mathrm i})\left[\sum _{j=0}^N T_j^2 P_j(f) + 2\sum _{j < m} (-1)^{j+m} C_{jm}(f) T_j T_ m \cos(\alpha \Delta_{jm})\right].
 \label{eq:XRS}
\end{equation}
In Eq.~\ref{eq:XRS} the layers are numbered from the substrate toward the surface. The spatial frequency $f$ is still related to the scattering and the incidence angles via Eq.~\ref{eq:grating}, $\alpha = 2\pi(\sin \theta_{\mathrm s}+\sin\theta_{\mathrm i})/\lambda$, $T_j$ is the field amplitude transmittance in the $j^{\mathrm {th}}$ layer, $\Delta_{jm}= \langle z_j \rangle - \langle z_m\rangle$ is the average distance between the $j^{\mathrm {th}}$ and the $m^{\mathrm {th}}$ interface, and 
\begin{equation}
	K(\lambda, \theta_{\mathrm s}, \theta_{\mathrm i}) = \frac{16\pi^2}{\lambda^3} Q_{\mathrm{is}}\,\sin^2\theta_{\mathrm s} \sin\theta_{\mathrm i}
	\label{eq:propfact}
\end{equation}
is the same proportionality factor appearing in Eq.~\ref{eq:singleXRS}. The polarization factor is still provided by Eq.~\ref{eq:polarization}, being $R_{\mathrm F}$ the intensity reflectivity of the single boundary between layers. Beyond the critical angles the values of $R_{\mathrm F}$ depend only a little on whether the reflection occur at the high-to-low density transition or the low-to-high one.   

It should be noticed that in Eqs.~\ref{eq:XRS} and \ref{eq:propfact} both incidence and scattering angles are corrected for refraction in the stack. This is possible only if the incidence angle is beyond the critical one of both materials, otherwise the refraction angle will no longer be meaningful. Therefore, to maximize the intensity of the scattered beam, it is convenient to set the incidence angle at one of the first peaks after the critical angle of the multilayer. Secondly, it should be kept in mind that Eq.~\ref{eq:XRS} becomes inaccurate at scattering angles smaller then the critical one. 

An important point is the computation of the $T_j$ coefficients. In a previous paper~\cite{Canestrari 2006} we assumed an exponential decrease of the intensity throughout the stack \cite{Spiga 2005} when rays impinge at a Bragg angle. The method hereby adopted is applicable to both periodic and graded multilayers and makes use of the recursive theory of multilayer reflectivity to derive the transmittance of the partial stack of the outermost $N-j$ layers. We have validated the recursive method by comparing its results with the findings of the IMD program \cite{Windt 1998}, finding a very good agreement.

The result of Eq.~\ref{eq:XRS} was added to a gaussian function at $\theta_{\mathrm s}=\theta_{\mathrm i}$ to simulate the specular reflected beam. Finally, the simulated XRS curve has been normalized to the flux impinging the sample, $I_0$, and smoothed to account for the finite detector acceptance angle $\mbox{d} \theta_{\mathrm s}$. 

The simulated scans are compared to the experimental XRS curves, obtained with the apparatus described in the next section. 

\section{Experimental}\label{sec:experimental}
\subsection{Samples}\label{sec:samples}
In this work we consider two samples as test cases:
\begin{enumerate}
	\item{a periodic multilayer with 40 bilayers of W/Si with $d_{\mathrm W}$ = 19.3 \AA~and $d_{\mathrm Si}$ = 26.7 \AA~deposited by e-beam evaporation onto a Silicon wafer (already characterized and analyzed \cite{Canestrari 2006}) and}
	\item{a graded multilayer with 100 bilayers of Pt/C deposited by magnetron sputtering onto a Silicon wafer, with layer thickness values as per the widespread supermirror design \cite{Joensen 1995}, given by the power law $d(j)=a(b+j)^{-c}$ 
	where $d$ is the d-spacing and $j$ is the bilayer index from the top of the stack. The nominal power-law parameter values are
	\begin{tabbing}
	Pt : \= $a$ = 31.0 \AA, \= $b$ = -0.94, \= $c$ = 0.23, \\
	C: \> $a$ = 53.0 \AA, \> $b$ = -0.88, \> $c$ = 0.21\\
	\end{tabbing}
being Platinum the first layer deposited onto the substrate. The outermost layer is Carbon.} 
\end{enumerate}
\subsection{Atomic Force Microscope}\label{sec:AFM}
The surface topography of the samples before and after coating was measured with the stand-alone Atomic Force Microscope (AFM) operated at INAF/OAB, a Veeco Instruments, mod. Explorer\textsuperscript{\textregistered}. The scanner can image surface areas of 1$\mu$m to 100 $\mu$m side. The AFM was operated in tapping mode, using an Antimony-doped Silicon probe with a proper frequency in the 100-250 kHz range. The maximum lateral resolution is 4 nm and the vertical resolution is better then 1~\AA.

Several scans were performed on each sample to check the uniformity of the deposition process and the possible presence of different kinds of defects. From 10~$\mu$m and 1~$\mu$m wide scans we computed the PSDs of the substrate and the outer surface of the multilayer in the spatial wavelength range 8~nm --10~$\mu$m. This spectral range is also typically involved in the roughening process \cite{Canestrari 2006}: at larger wavelengths, the rough profiles are essentially replicated without relevant roughness increase, whilst high frequencies are damped out.

\subsection{X-ray Reflectivity}\label{sec:XRR}
The thickness values of the multilayers under test were investigated via X-ray Reflectivity (XRR) measurements with the BEDE-D1 diffractometer from Bede Scientific Instruments Ltd\textsuperscript{\textregistered}. A Cu anode X-ray tube is used as source. A channel-cut Si crystal monochromator and a slit collimator are used to filter and shape the Cu K$\alpha$1 fluorescence line at 8.045~keV: the first slit (50 $\mu$m wide) removes the X-ray continuum and the K$\alpha$2 line diffracted by the monochromator, whereas the second slit (10 $\mu$m wide, very close to the sample) reduces the beam width for the sample to collect it entirely, also at very small incidence angles. The reflected beam is then collected through a 800~$\mu$m wide slit by a scintillation detector with a pulse-height discriminator that limits the intrinsic background to $<$ 0.2~counts/sec. The measurement is performed via a $\theta-2\theta$ scan with 20 arcsec steps, on the order of the achieved beam divergence (15 arcsec). After the scan, the reflectance is normalized to the incident flux. The detailed fit of the XRR curve, to determine the stack parameters (thickness, density, roughness), is performed using the PPM program \cite{Spiga 2007}. 

\begin{figure}[!tpb]
        \centering
        \includegraphics[width=\textwidth]{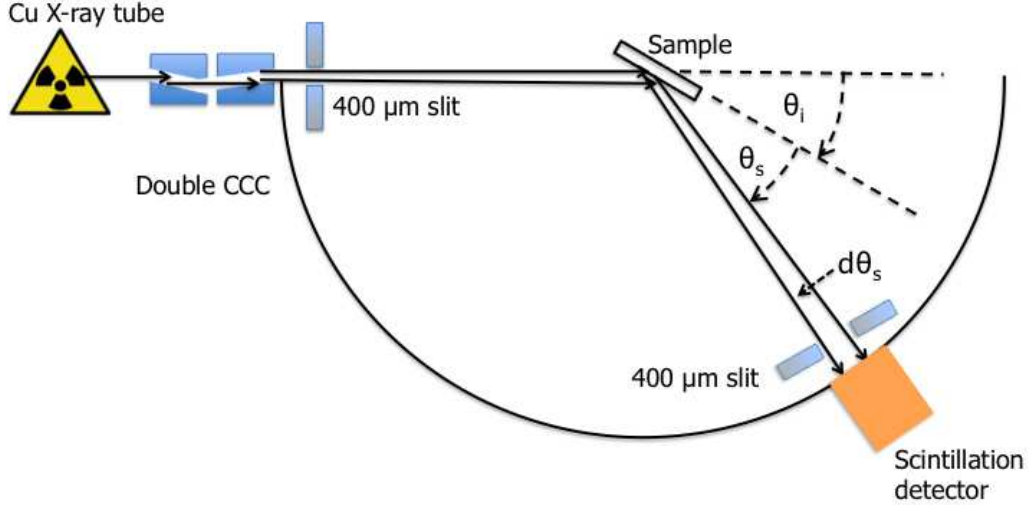}
        \caption{Scheme of the XRS experimental equipment.}
        \label{fig:schemadiff}
\end{figure}

\subsection{X-ray Scattering}\label{sec:XRS}
The BEDE-D1 diffractometer was also used to measure the scattering of 8.045 keV (1.541 \AA) X-rays at incidence angles corresponding to the first peaks after the critical angle. In this case, a double channel-cut Si crystal monochromator was used in order to improve the beam collimation (to within 10~arcsec FWHM). In this configuration, a thin slit is not needed to isolate the Cu K$\alpha$1 line. On the other hand, the incoming flux has to be intense in order to scan the scattered beam up to angles as large as 15000~arcsec. Moreover, the incidence angle will be near 1~deg from the surface; therefore, the beam does not need to be very thin to be entirely collected on the sample, so a 400~$\mu$m-wide slit (wider than the one used for the XRR) was positioned after the monochromator. A 400~$\mu$m slit in front of the detector, at a 340~mm distance from the sample, provides a sufficiently small angular acceptance for our scopes ($\mbox{d} \theta_{\mathrm s} \approx $ 320~arcsec FWHM). The measurement of the scattered intensity, $\mbox{d} I_{\mathrm s}$ (Eq.~\ref{eq:XRS}), was performed by scanning the detector in the plane of incidence with 20~arcsec steps. The layout of the experimental apparatus is sketched in Fig.~\ref{fig:schemadiff}.

Application of Eq.~\ref{eq:grating} shows that a detector scan with $\theta_{\mathrm s} \in [\sim 3000 \div 15000]$~arcsec, with $\theta_{\mathrm i} $ = 2200~deg, at $\lambda$ = 1.541~\AA, corresponds to spatial wavelengths between a few microns and fractions of micron, i.e., the ones in the sensitivity window of the AFM (Sect.~\ref{sec:AFM}) and mostly involved in the roughness growth process.

\section{Results vs. modeling}\label{sec:results}
\subsection{Sample A}\label{sec:sampleA}
\begin{figure}[!htpb]
	\begin{subfigure}{0.5\textwidth}
        		   \centering
       		   \includegraphics[width=\textwidth]{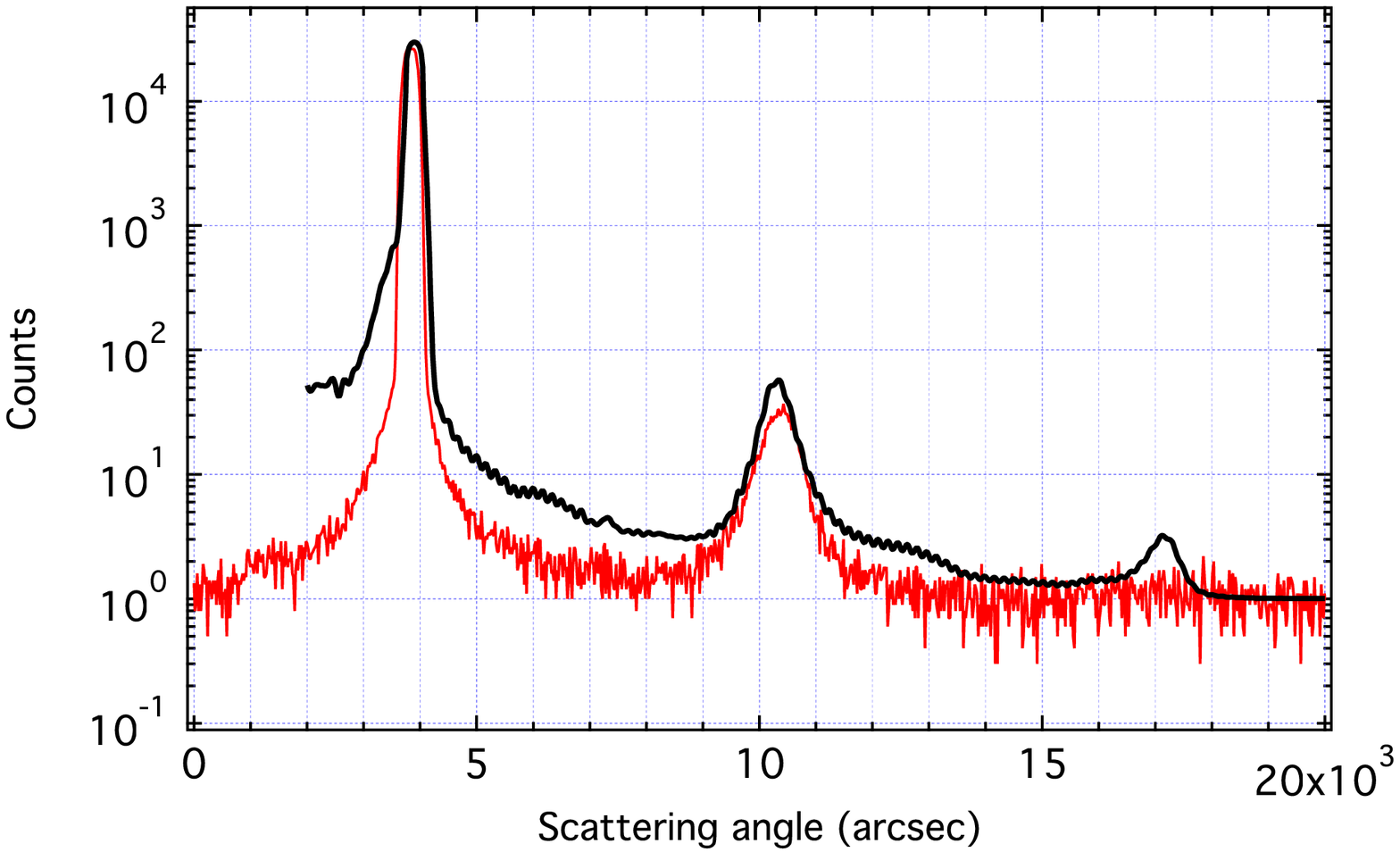}
     	            \caption{}
    	            \label{fig:periodico_nogrowth_fields}
         \end{subfigure}
         	\begin{subfigure}{0.5\textwidth}
         		\centering
         		\includegraphics[width=\textwidth]{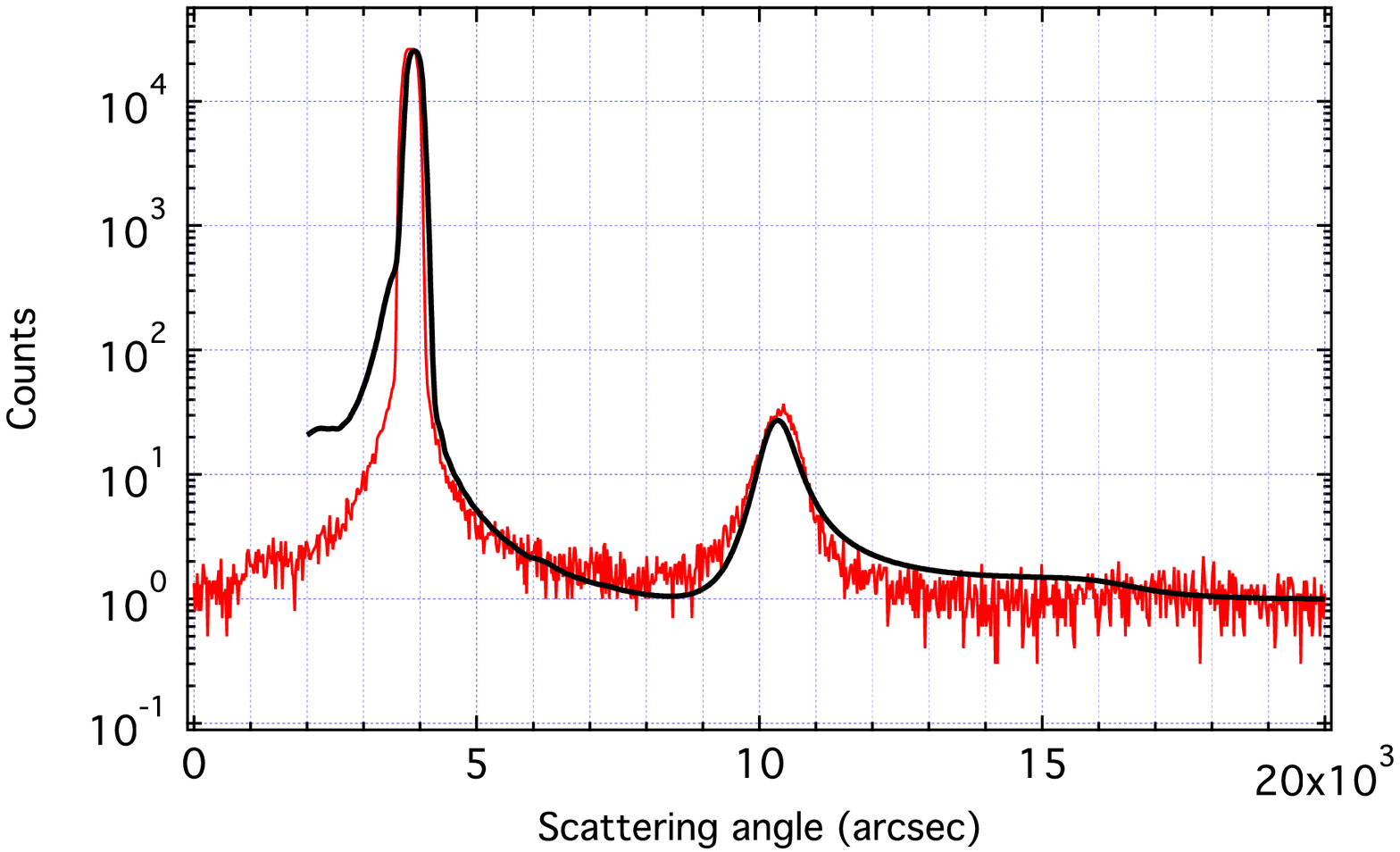}
         		\caption{}
      	   	\label{fig:periodico_growth_fields}
         \end{subfigure}
         \caption{Experimental (red lines) X-ray scattering for the sample A vs. modeling (black lines). (a) Assuming a constant PSD in the stack returns a poor modeling. (b) Accounting for the roughness evolution through the stack reproduces more correctly the experimental XRS curve.}
         \label{fig:periodico_fields}
\end{figure}
The complete characterization of this sample was already reported in a previous work \cite{Canestrari 2006}. The roughness growth was also studied in detail therein. The XRS scan, taken at the $1^{\mathrm{st}}$ Bragg angle of incidence, exhibits an apparent peak at $\theta_{\mathrm s} \sim$ 10000 arcsec, stemming from the constructive interference of scattered waves. In \cite{Canestrari 2006}, however, the computation was performed using an exponential trend for the $T_j$ coefficients (Sect.~\ref{sec:X-model}), which can be only applied to periodic multilayers in Bragg incidence. We have reconsidered those data to validate the general method used heretofore to compute the $T_j$'s. The experimental XRS scan vs. the result of the modeling out of Eq.~\ref{eq:XRS} is displayed in Fig.~\ref{fig:periodico_fields}. In order to show the impact of the PSD evolution in the XRS diagram, we have preliminarily assumed no evolution of the PSD throughout the stack with a complete correlation ($C_{jm} = P_j = P_{\mathrm{outer}}$) at all frequencies. The modeling clearly overestimates the measurement (Fig.~\ref{fig:periodico_nogrowth_fields}): as expected, this denotes a gradual evolution of the roughness from the substrate from the outer surface. Moreover, the hypothesis of a complete correlation at all frequencies ($a_j =1$ for all $j$, Eq.~\ref{eq:replication})  endows the simulation with a second XRS peak near 17000~arcsec, which is not observed in the experimental XRS curve. In contrast, setting the Spectral Density functions resulting from the correct modeling of the roughness growth yields a modeling in much better accord with the experimental scan (Fig.~\ref{fig:periodico_growth_fields}). The model-data matching also proves the correct trend of the electric field coefficients throughout the stack, computed using the general method.

\subsection{Sample B}\label{sec:sampleB}
\begin{figure}[!htpb]
	\begin{subfigure}{0.5\textwidth}
        		   \centering
      		   \includegraphics[width=\textwidth]{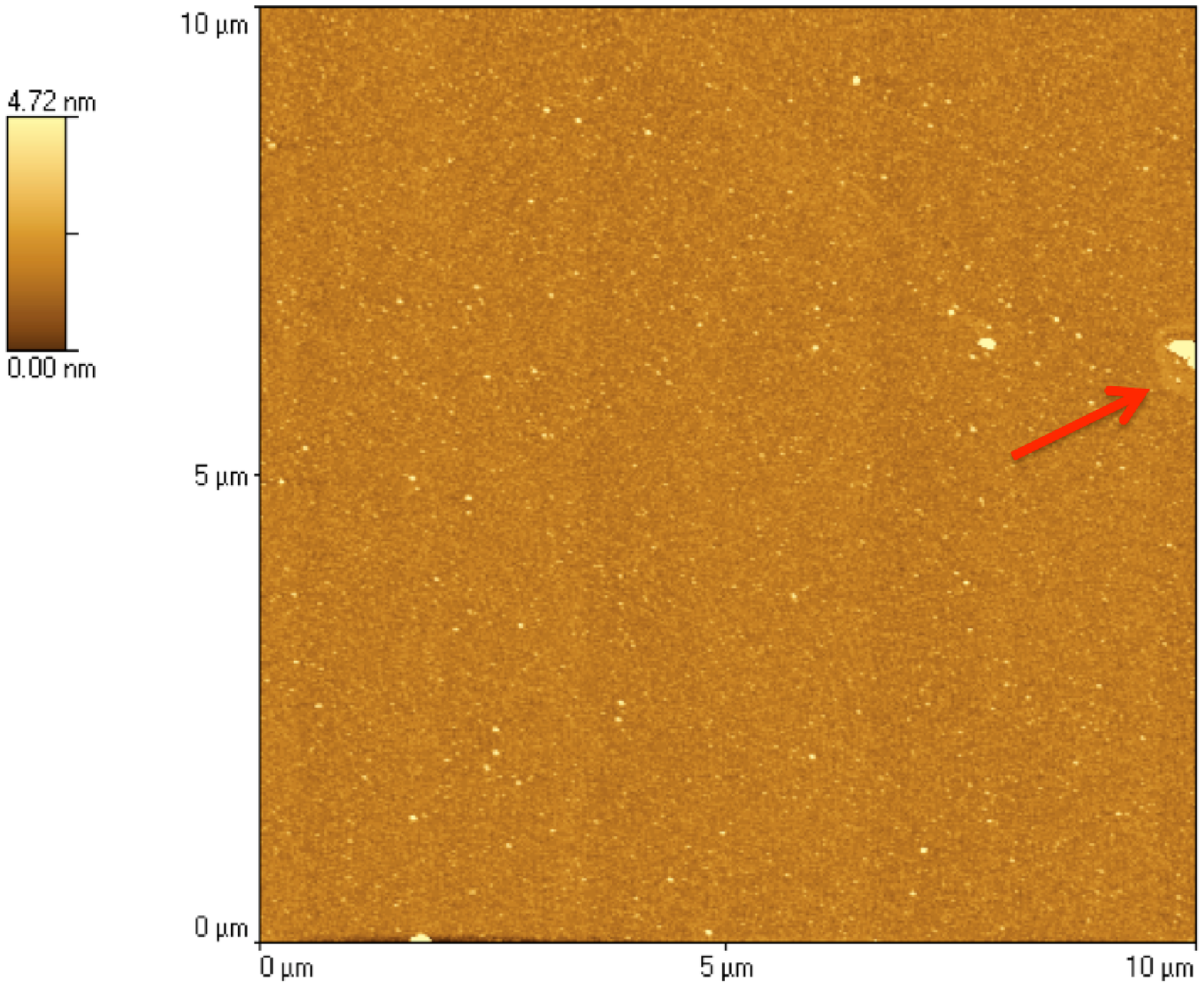}
     	            \caption{}          
    	            \label{fig:afm}
	\end{subfigure}
         	\begin{subfigure}{0.5\textwidth}
	  \centering
      		   \includegraphics[width=\textwidth]{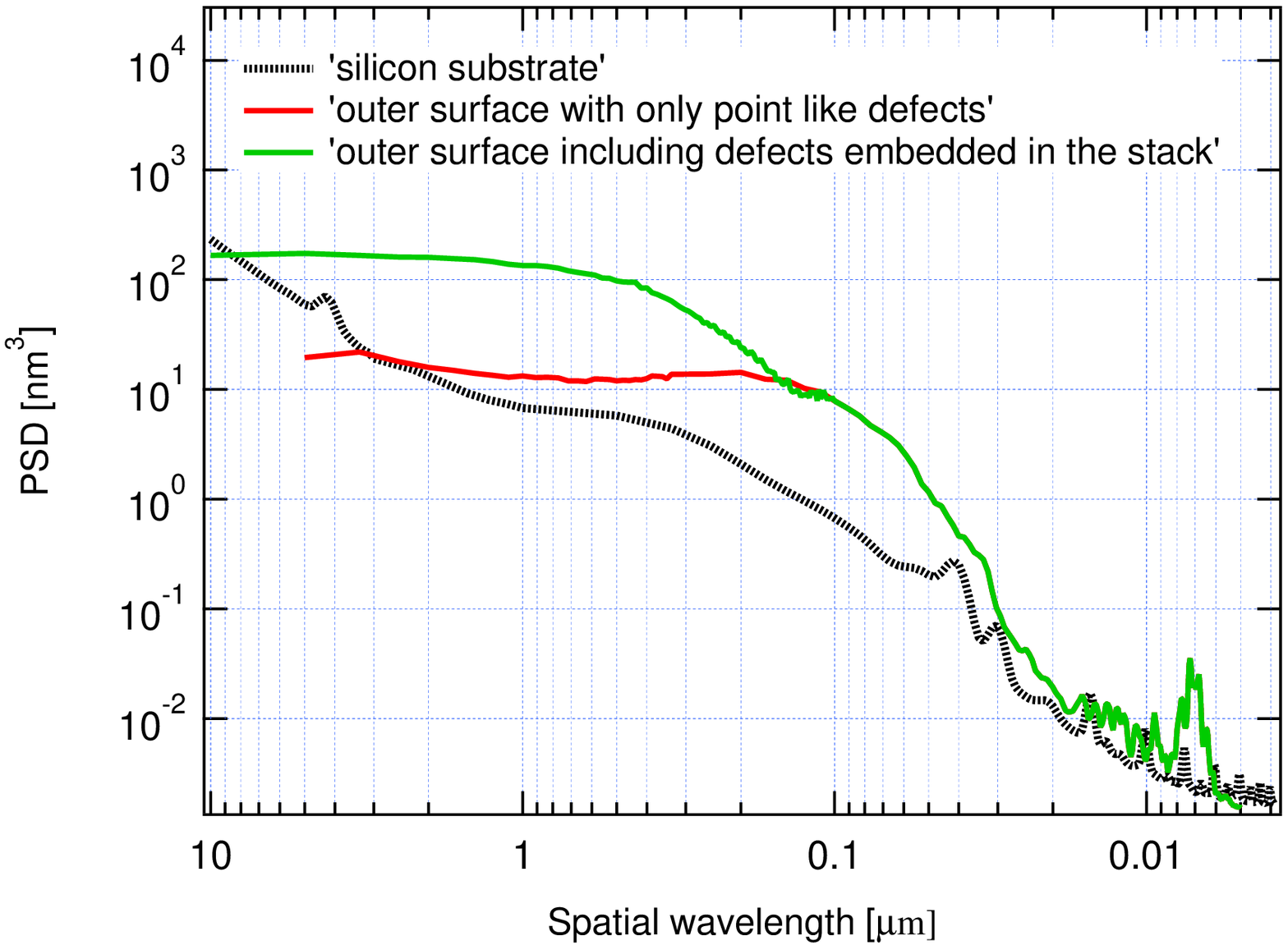}
     	            \caption{}
    	            \label{fig:PSD}
	\end{subfigure}
	\begin{subfigure}{0.5\textwidth}
        		   \centering
      		   \includegraphics[width=\textwidth]{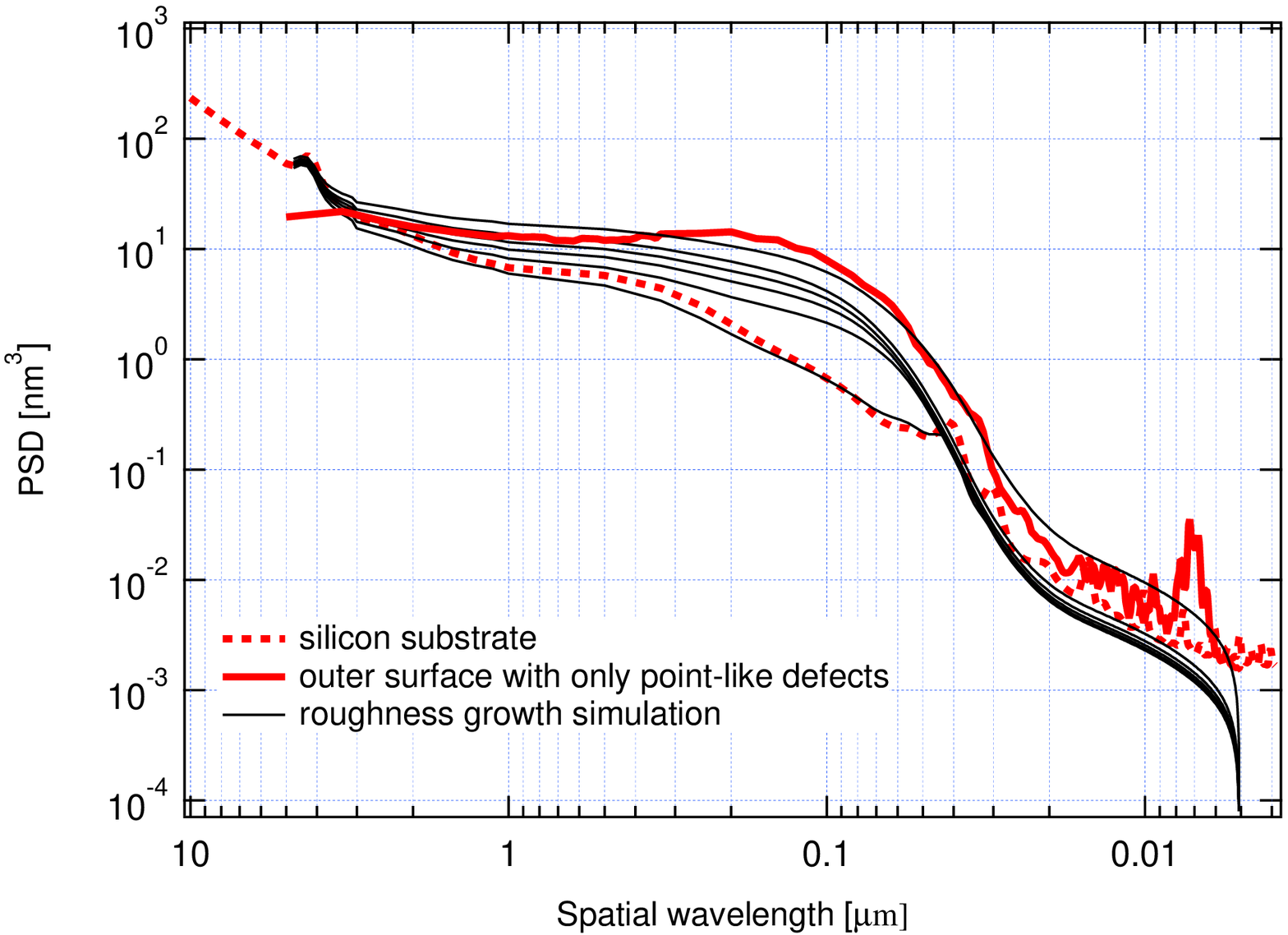}
     	            \caption{}          
    	            \label{fig:growth_pointlikedefects}
	\end{subfigure} 
         	\begin{subfigure}{0.5\textwidth}
	  \centering
      		   \includegraphics[width=\textwidth]{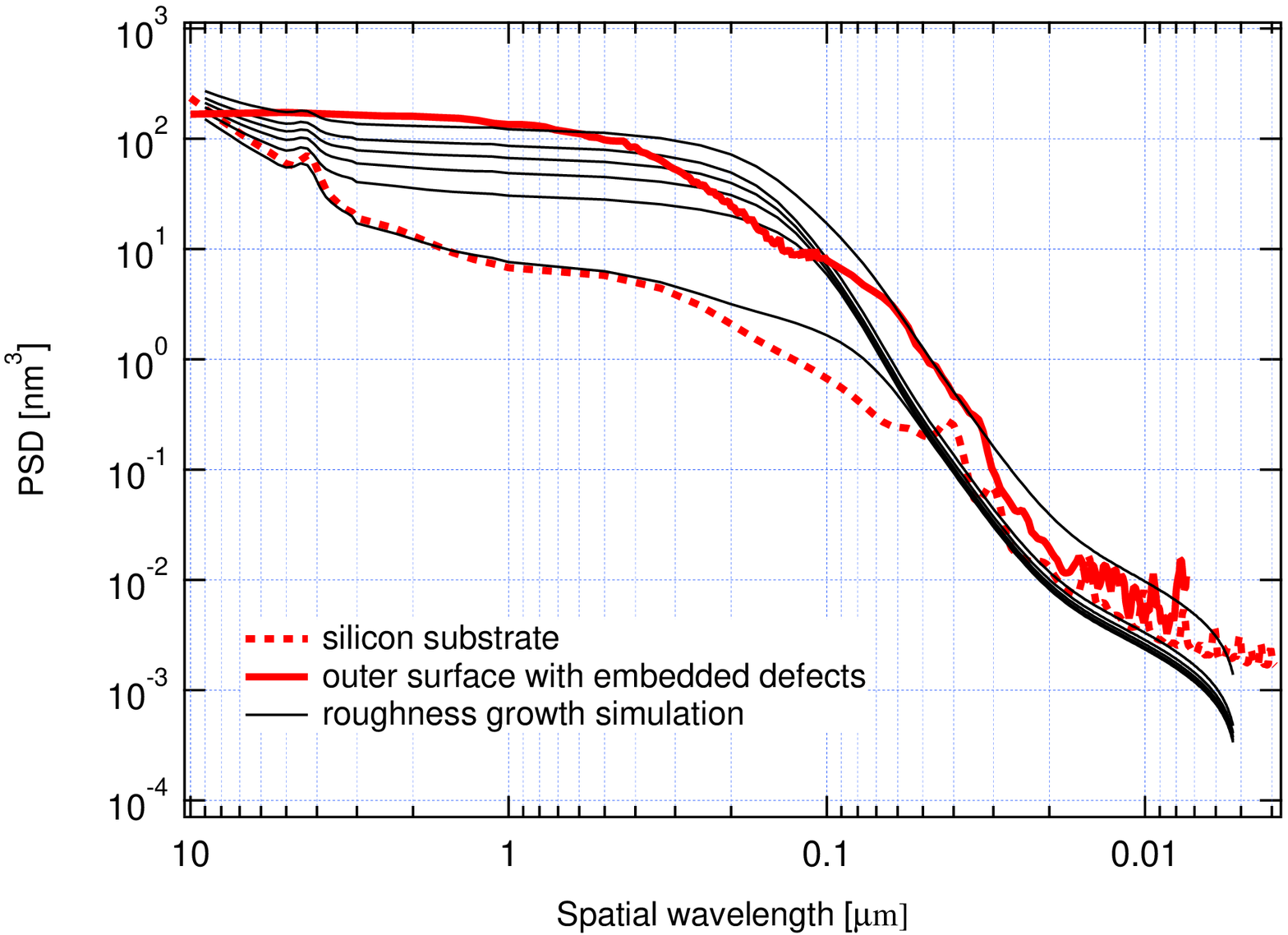}
     	            \caption{}
    	            \label{fig:growth_embedded}
	\end{subfigure}
	   \caption{Roughness analysis from AFM scans of the outer surface of the sample B. The AFM image (a) shows that the surface is crowded with point-like defects in ejection, and also exhibits some defects of bigger size. (b) The bigger defects can be included or excluded from the PSD computation. The PSD evolution in the stack was consequently computed (c) with and (d) without the major defects. The XRS measurement shows (Fig.~\ref{fig:XRS-R5fitB}) that the correct modeling is (d).}
\label{fig:afm_PSD}
\end{figure}

For this graded multilayer sample we have directly measured with the AFM (Sect.~\ref{sec:AFM}) the roughness of the outermost Carbon layer in the 10~$\mu$m -- 5~nm spatial wavelength range, while the substrate roughness was supposed to be the same of a standard Silicon wafer \cite{Vernani 2006}, e.g., 2 \AA~rms in that spectral window. The multilayer surface, as measured with the AFM, exhibits crowded point-like defects in ejection (Fig.~\ref{fig:afm}) that increase the rms of the external surface to 3.7~\AA~in the same spectral range. Besides, other defects of bigger size increase the rms from 3.7 \AA~to 7.3 \AA~in the 10 $\mu$m scan (Fig.~\ref{fig:PSD}). These bigger defects are homogeneously distributed over the sample, but it is difficult to ascertain if they are surface contaminations or they stem from the roughness growth itself. In the latter case, such defects would be effective for X-ray scattering, whilst in the former one, 8.045 keV X-rays would be almost unaffected by their presence. For this reason, we have performed two modelings of the roughness growth, including or excluding the biggest defects. The two PSD evolutions are shown in Fig.~\ref{fig:growth_pointlikedefects} and~\ref{fig:growth_embedded}. Both models were used to simulate the expected XRS diagram: the comparison to the experimental curve will show that the modeling of Fig.~\ref{fig:growth_embedded} is correct.

\begin{figure}[!htpb]
       	\centering
      	\includegraphics[width=0.7\textwidth]{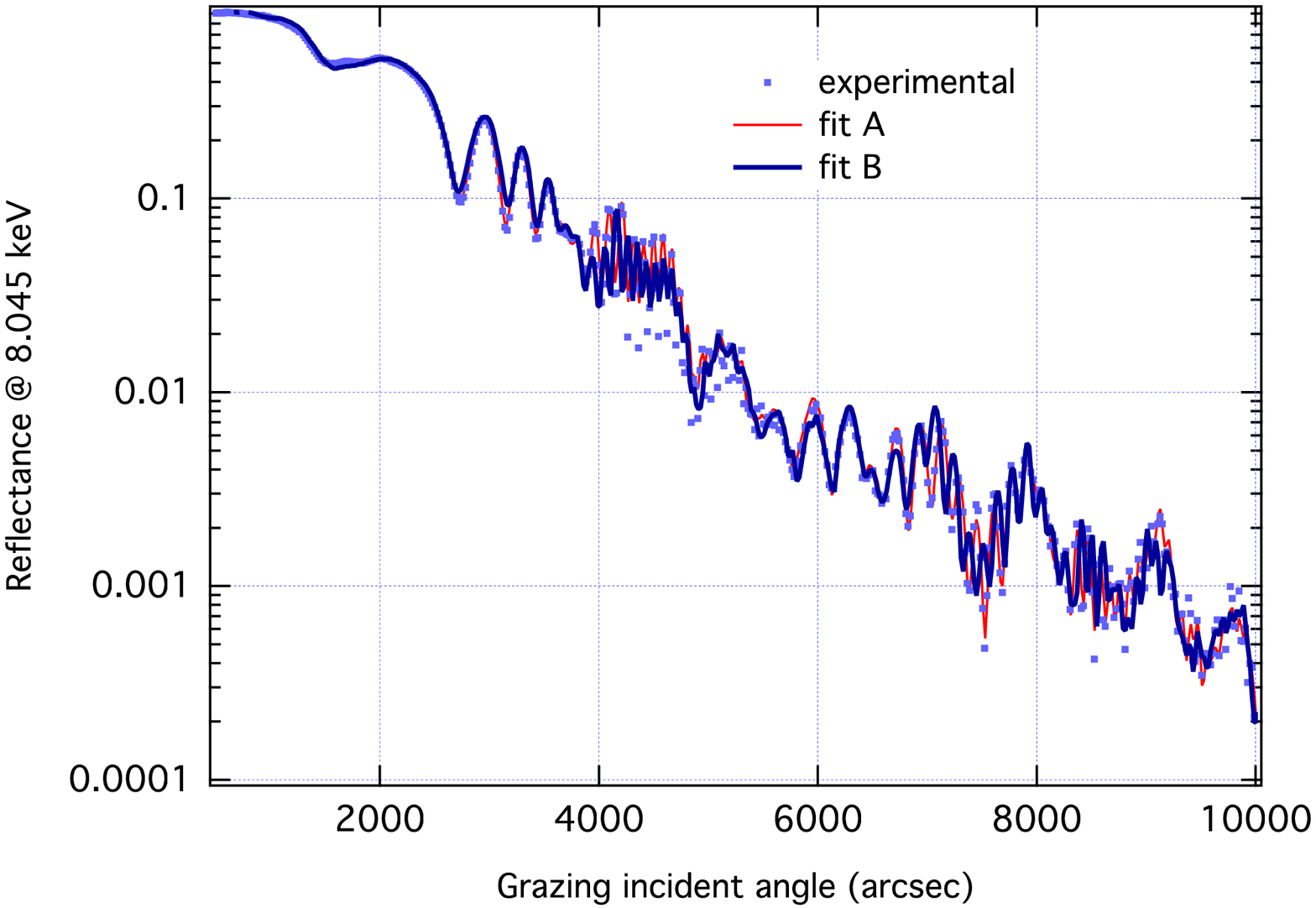}
	 \caption{XRR angular scan at 8.045 keV of sample B, compared to the reflectivity computed from two slightly different stack models. They both match the XRR measurement, but only the fit 'B' is also in agreement with the peak positions seen in the XRS scans (Fig.~\ref{fig:XRSfitB}).}
	\label{fig:XRR}
\end{figure}

\begin{figure}[!htpb]
	\begin{subfigure}{0.5\textwidth}
        		   \centering
      		   \includegraphics[width=7cm]{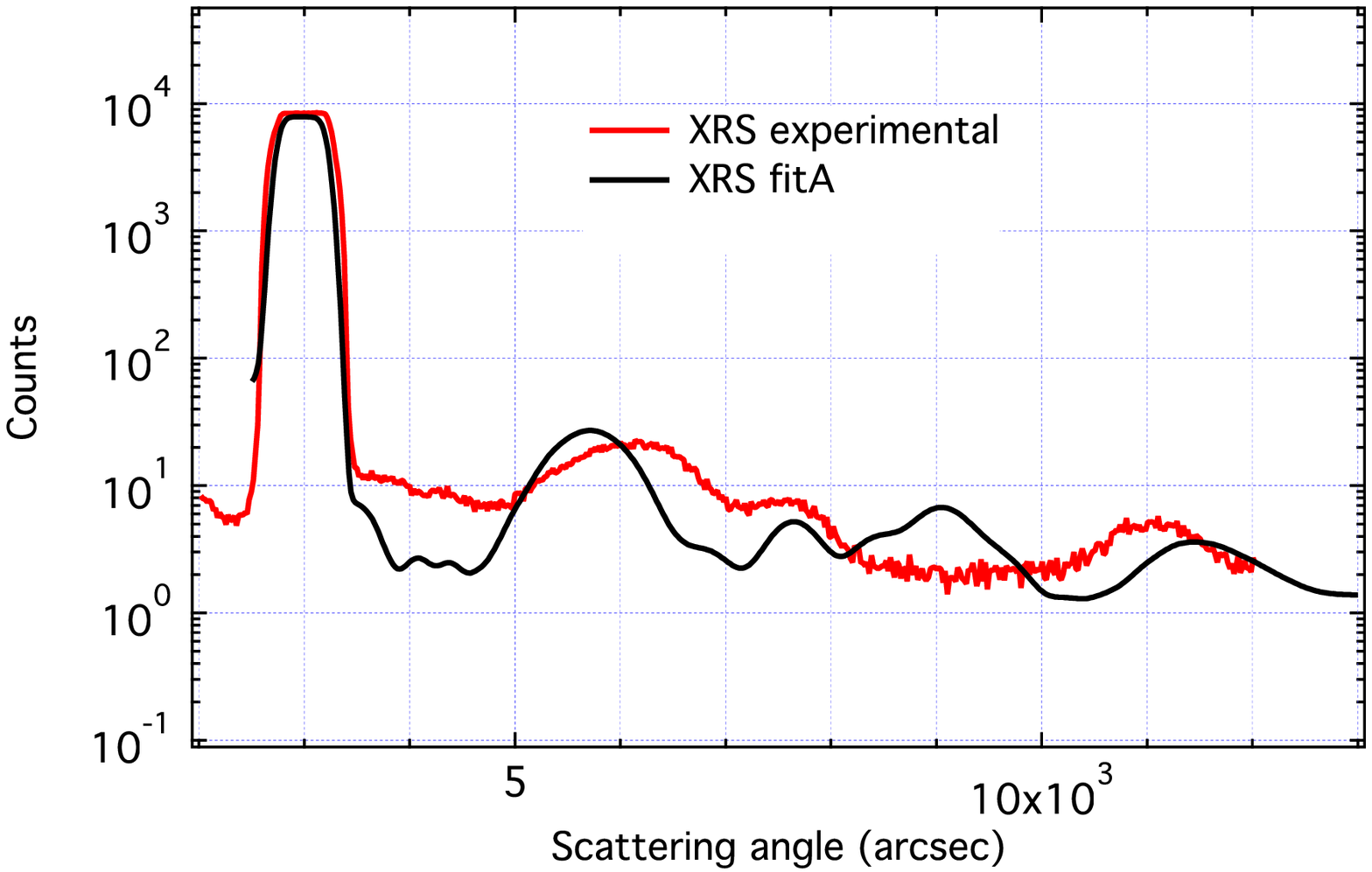}
	            \caption{}
	            \label{fig:XRSfitA}
	\end{subfigure} 
         \begin{subfigure}{0.5\textwidth}
        		   \centering
      		   \includegraphics[width=7cm]{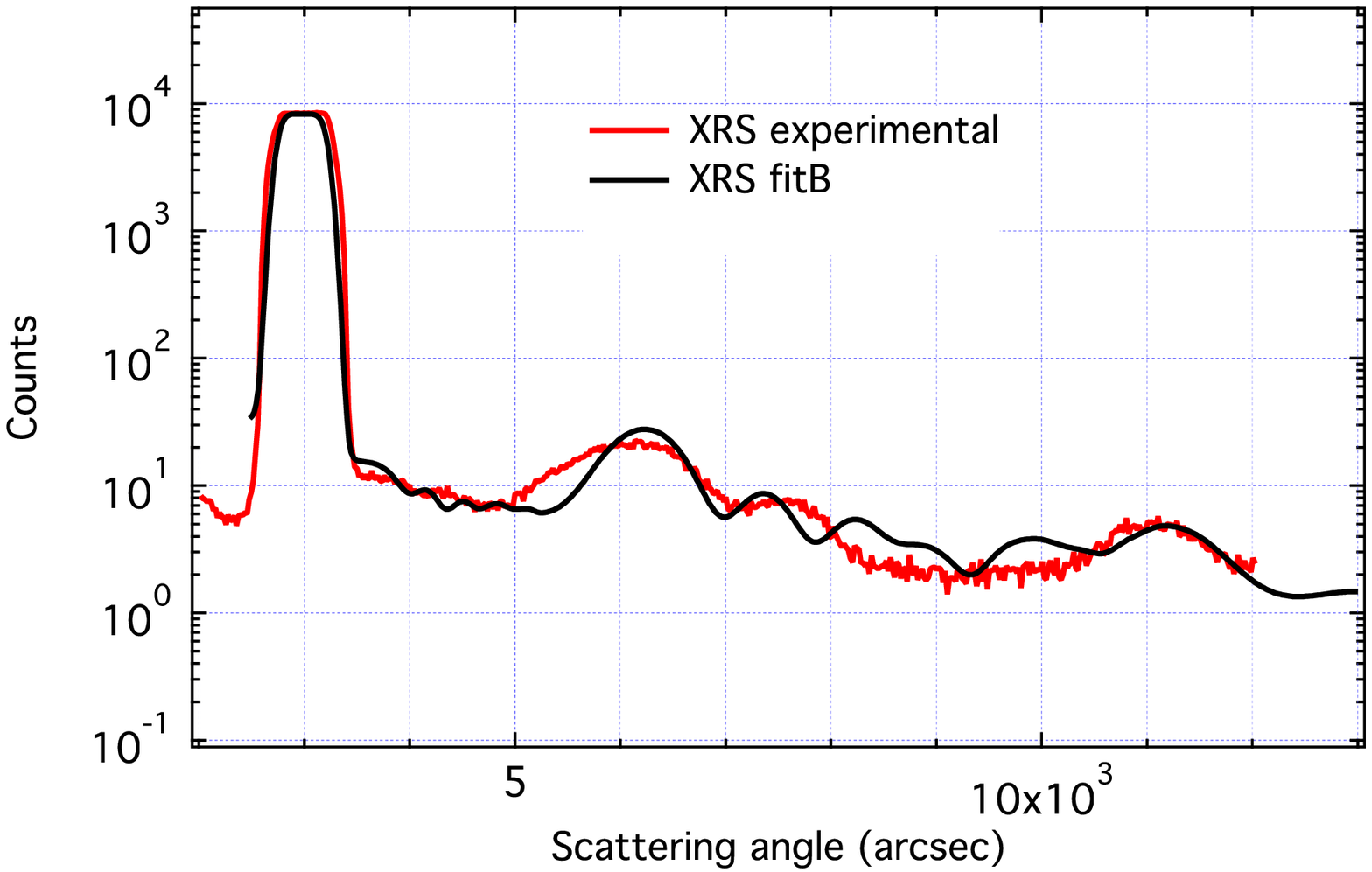}
     	            \caption{}          
    	            \label{fig:XRSfitB}
	\end{subfigure} 
\caption{Experimental XRS of the sample B (red lines) at $\theta_{\mathrm i}$ = 3000~arcsec, compared with the simulated ones from the PSD evolution, including the major defects (Fig.~\ref{fig:growth_embedded}: black lines). (a) XRR fit 'A', (b) XRR fit 'B'. The layer thickness values provided by the fit 'B' of Tab.~\ref{table:powerlaw} reproduces much better the observed peak positions.}
	\label{fig:XRS2011fitAB}
\end{figure}

Prior to the scattering measurement, the XRR of the sample at 8.045 keV has been measured (Sect.~\ref{sec:XRR}). The PPM program was subsequently applied to perform a detailed fit of the reflectivity \cite{Spiga 2007} and return the best-fit power-law parameters for the stack (Tab.~\ref{table:powerlaw}, fit 'A'). The layer thickness values were used to firstly model the roughness growth (Eq.~\ref{eq:Pmultilayer}) and to subsequently simulate the XRS diagram (Eq.~\ref{eq:XRS}) of X-rays impinging at the second peak after the critical angle ($\theta_{\mathrm i} \simeq$ 3000~arcsec). The electric fields were computed using the recursive method, which is the only one applicable for a graded multilayer and already tested with the Sample A (Sect.~\ref{sec:sampleA}). We already note that adopting the PSD evolution modeling that includes all the surface defects (Fig.~\ref{fig:growth_embedded}) yields the correct scattering level. Nevertheless, the XRS diagram simulated in this way did not perfectly fit the measured XRS (Fig.~\ref{fig:XRSfitA}); for instance, the peak positions did not match. This denotes some departure of the actual thickness trend from the modeled one.

We have so found that a slightly different power law (Tab.~\ref{table:powerlaw}, fitB) matches much better the scattering peak positions (Fig.~\ref{fig:XRSfitB}). This new power-law actually returns layer thickness values that do not differ by more than 3 \AA~ from the first one, but in the outermost C layer, which turns out to be 15 \AA~thicker. Quite surprisingly, the XRR measurements actually matches both A and B power laws (Fig.~\ref{fig:XRR}). A possible interpretation could be that the specular reflectivity at 8.045 keV is a little sensitive to the thickness of the outer C layer, because most of the XRR curve features that drive the fit program are determined more by the multilayer internal structure, rather than by the thickness of the capping layer. In contrast, an XRS measurement can be more sensitive to the increased roughness generated by a thicker C layer. The experimental scattering diagram is now well reproduced by the modeling ($\chi^2_{\nu} \approx $ 0.15). We therefore assumed the power law 'B' of Tab.~\ref{table:powerlaw} to be the correct one.
\begin{table}[!htpb]
\caption{Fitting power law parameters} 
\centering 
\begin{tabular}{c c c } 
\hline\hline 
Case & Pt: a [\AA], b, c & C: a [\AA], b, c  \\ [0.5ex] 
\hline 
Fit A & 31.08, -0.90, 0.30 & 53.16, -0.96, 0.19 \\ 
Fit B & 31.00, -0.94, 0.23 & 53.00, -0.88, 0.21  \\ [1ex] 
\hline 
\end{tabular}
\label{table:powerlaw} 
\end{table}

\begin{figure}[!htpb]
	\begin{subfigure}{0.5\textwidth}
        		   \centering
      		   \includegraphics[width=7cm]{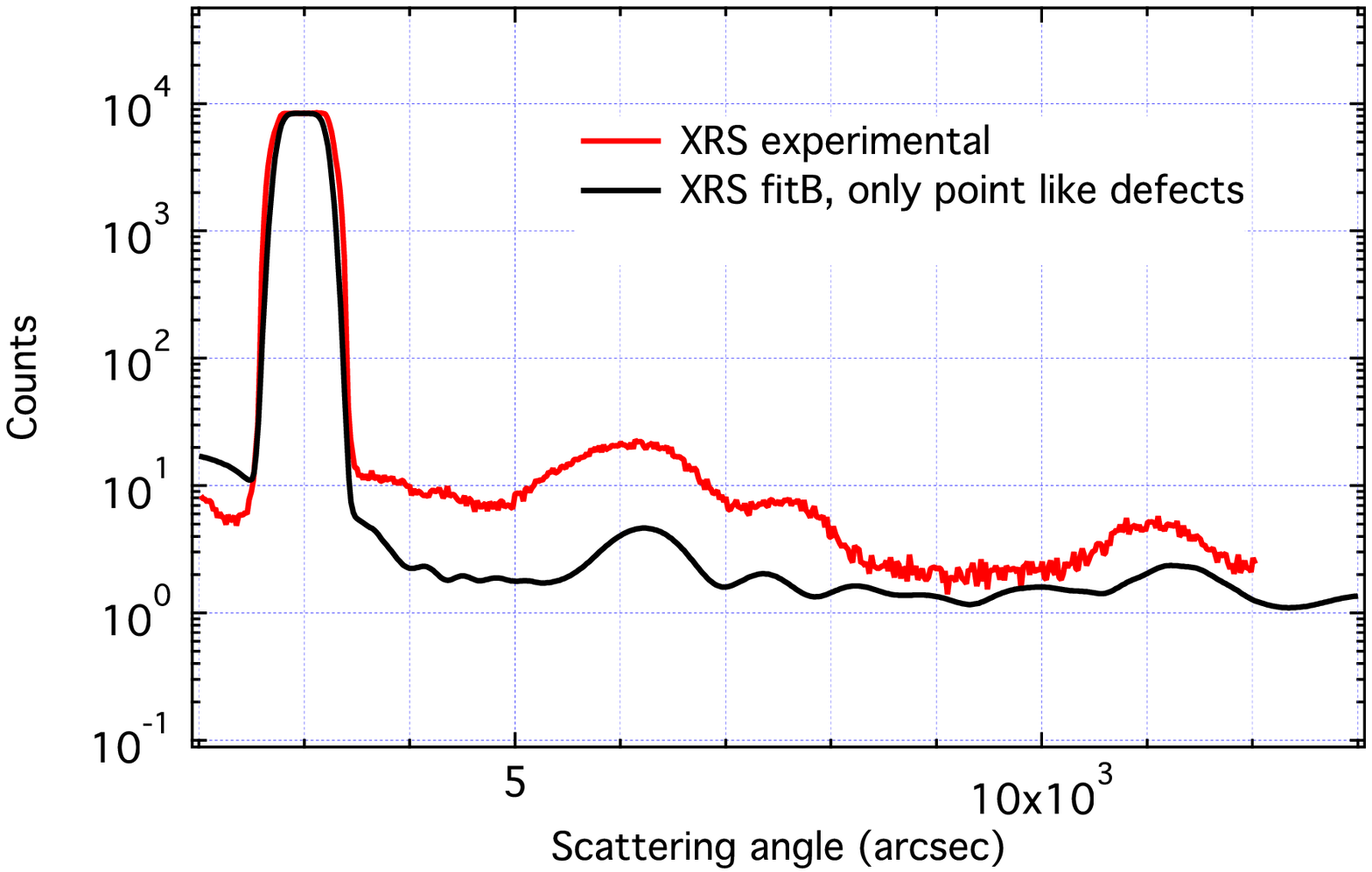}
     	            \caption{}          
    	            \label{fig:XRS-fitBnodefects}
	\end{subfigure}
	 \begin{subfigure}{0.5\textwidth}
        		   \centering
      		   \includegraphics[width=7cm]{XRS-2982-2011-fitB.eps}
     	            \caption{}          
    	            \label{fig:XRS-R5fitB}
	\end{subfigure}
\caption{Experimental XRS for sample B (red lines) at $\theta_{\mathrm i}$ = 3000~arcsec, assuming the fit 'B' to model the stack structure, compared with the simulated ones from the PSD evolution (black lines): (a) considering only the point-like defects of Fig.~\ref{fig:afm}, (b) considering all the visual defects in the AFM maps.}
	\label{fig:XRS-2011-nomatch}
\end{figure}

Modeling the stack with the parameters given by the fit B, we hereafter consider the effects of including different defects in the PSD, and therefore in the XRS computation. If the sole point-like defects are included (Fig.~\ref{fig:afm}) and the growth parameters are tuned accordingly, the peak heights are underestimated in the simulated XRS diagram (Fig.~\ref{fig:XRS-fitBnodefects}). Conversely, if all the visual defects in the AFM maps are used to compute the final PSD of the multilayer, the scattering diagram matches much better the experimental XRS curve (Fig.~\ref{fig:XRS-R5fitB}). We therefore conclude that the visual defects other than the point-like defects observed on the multilayer surface (Fig.~\ref{fig:afm}) are related to the deposition process, and not just dust contaminations.

\begin{figure}[!htpb]
         \centering
         \includegraphics[width=0.7\textwidth]{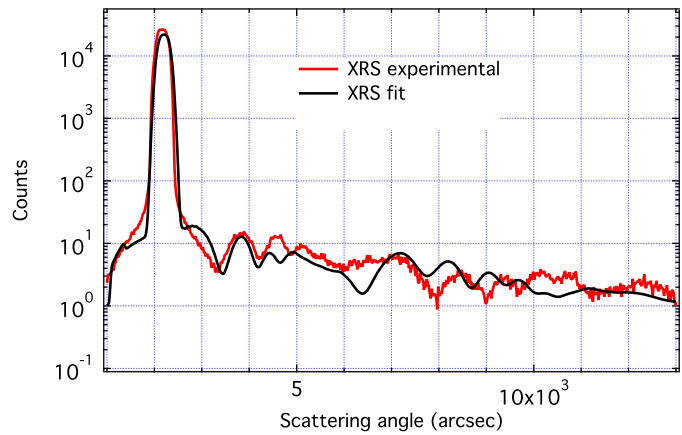}
	\caption{Experimental XRS for sample B (red lines) at $\theta_{\mathrm i}$ = 2200~arcsec, assuming the fit 'B' to model the stack structure, compared with the simulated ones from the PSD evolution (black lines) including all the visual defects in the AFM maps.\label{fig:2200-2012}}
\end{figure}

A further validation has been obtained from an XRS measurement at a different incidence angle, i.e., the first XRR peak after the critical angle ($\theta_{\mathrm i}$ = 2200~arcsec). The comparison between theory and experiment (Fig.~\ref{fig:2200-2012}) shows a good accord also in this case ($\chi^2_{\nu} \approx $ 0.29), even if the peak positions are not exactly reproduced: the stack structure might be slightly inhomogeneous and the two XRS measurements, performed in two different runs, might have been performed at different locations. Consequently, also the power law parameters might require some adjustment. 

\section{Conclusions}\label{sec:summary}
The XRS sensitivity of a multilayer to the roughness evolution and to the actual thickness trend in the stack makes it a powerful diagnostic instrument, provided that it is coupled to an opportune modeling of the PSD evolution and that the XRS pattern can be reliably modeled (Eq. 4). So far \cite{Canestrari 2006, Schroder 2007} the modelization of XRS for multilayers was limited to the periodic case. In this paper we have shown that the formalism can be extended to graded multilayers. In the tested cases we have reached a good experiment-modeling simultaneous matching for both PSD growth and X-ray scattering. The comparison between XRR and XRS fitting shows that for graded multilayers the XRS is more sensitive than XRR to the actual thickness trend in the stack, since even a few angstr\"{o}ms variation significantly changes the XRS peak positions. Moreover, by fitting the experimental XRS, it becomes possible to discriminate defects that produce scattering from surface contaminations, an analysis impossible to obtain from the sole AFM scans. 

\section*{Acknowledgments}
This work was financed by the Italian Space Agency (contract I/069/09/0).

\appendix

\section{X-ray scattering in a multilayer coating}\label{appXRS}
\begin{figure}[!tpb]
        \centering
        \includegraphics[width=0.85\textwidth]{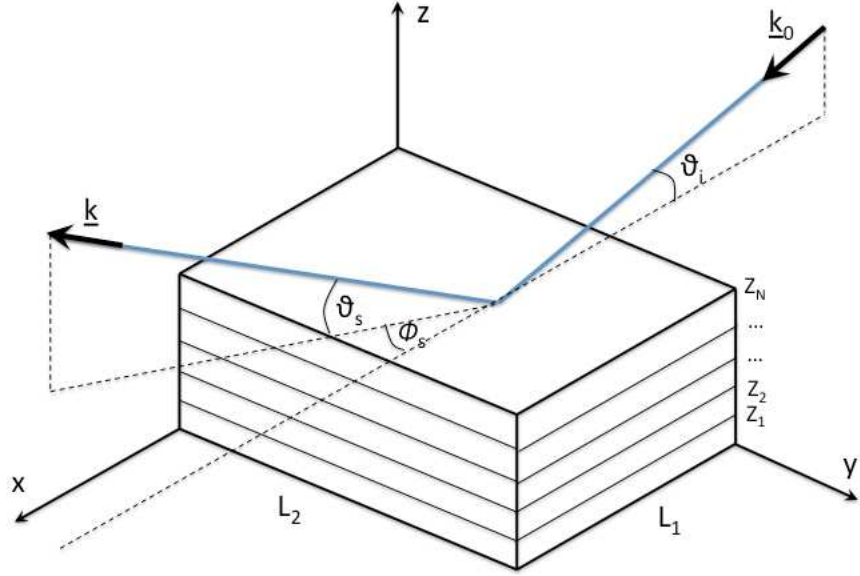}
        \caption{Reference frame used in the computation of the XRS diagram of a multilayer coating.}
        \label{fig:refframe}
\end{figure}

In this appendix we briefly derive Eq.~\ref{eq:XRS} used in Sect.~\ref{sec:X-model} to describe the X-ray scattering from a multilayer. The complete computation can be found in \cite{Spiga 2005}. Let the multilayer substrate surface to be a rectangle of sides $L_1$ and $L_2$ in the $xy$ plane (Fig.~\ref{fig:refframe}), with the coating that covers a surface $S$ at $z = 0$ and occupies a volume $V$ at $z >0$. The electric field of amplitude $E_0$ impinges the sample in the $xz$ plane at a shallow incidence angle $\theta_{\mathrm i}$ and partly travels across the coating. The electric field $E$($z$) amplitude varies with the depth in the stack owing to the progressive reflection and absorption in the multilayer. The scattered intensity results from the interference of elementary waves scattered by electrons in the entire stack. 

We denote with ${\cal N}(z)$ the number of free electrons per volume unit, as a function of the depth in the coating. This number is related to the material density $\rho$, the atomic weight $A$, the scattering coefficient $f_1$, as ${\cal N}(z) = \rho N_{\mathrm A}f_1/A$, where $N_{\mathrm A}$ is the Avogadro number. The scattered power $\mbox{d} I_{\mathrm s}$ into the solid angle $\mbox{d} \Omega_{\mathrm s} = \cos\theta_{\mathrm s}\mbox{d} \theta_{\mathrm s} \mbox{d} \phi_{\mathrm s} $ at the polar coordinates ($\theta_{\mathrm s}$, $\phi_{\mathrm s}$) can be expressed as:
\begin{equation}
	\frac {\mbox{d} I_{\mathrm s} }{\mbox{d} \Omega_{\mathrm s}} =\frac {c}{8\pi} \frac{\mbox{d}\sigma}{\mbox{d}\Omega_{\mathrm s}} \left| \int _V E(z)\, {\cal N}(z)\, e^{-i(\underline{k}-\underline{k}_0)\cdot\underline{r}}\,\mbox{d}^3\underline {r} \right|^2,
	\label{eq:intscatt1}
\end{equation}

where $c$ is the speed of light, $\underline{k}_0$ is wave vector of the incident wave, $\underline{k}$ is wave vector after the scattering, $\underline{r} = (x,y,z)$ is the location of the scattering electron, and $\mbox{d}\sigma/\mbox{d}\Omega_{\mathrm s} = \frac{1}{2}r^2_{\mathrm e}(1+\cos^2(\Delta\theta))$ is the Thomson cross section for unpolarized radiation, with $\Delta\theta$ the angular deviation and $r_{\mathrm e} = 2.8 \times 10^{-15}$~cm, the classical electron radius. Substituting the expressions of the wave vectors, we define
\begin{eqnarray}
	f_x &=& \frac{\cos\theta_{\mathrm s}\cos\phi_{\mathrm s}-\cos\theta_{\mathrm i}}{\lambda}, \label{eq:freqdef1}\\
	f_y &=& \frac{\cos\theta_{\mathrm s}\sin\phi_{\mathrm s}}{\lambda},\label{eq:freqdef2}\\
	\alpha &=& 2\pi\frac{\sin\theta_{\mathrm s}+\sin\theta_{\mathrm i}}{\lambda},\label{eq:freqdef3}
\end{eqnarray}
where the angles are evaluated at a proper average value of the refraction angles inside the stack. Using Eqs.~\ref{eq:freqdef1} through~\ref{eq:freqdef3}, we can rewrite Eq.~\ref{eq:intscatt1} as follows:
\begin{equation}
	\frac{\mbox{d} I_{\mathrm s} }{\mbox{d} \Omega_{\mathrm s}} =\frac {c}{8\pi} \frac{\mbox{d}\sigma}{\mbox{d}\Omega_{\mathrm s}} \left| \int_S \mbox{d}^2\underline{x}\, e^{-2 \pi i \underline{f} \cdot \underline{x}} \int _{-\infty}^{z_N}\!\! \mbox{d}z \, E(z) \,{\cal N}(z) \,e^{- i\alpha z} \right|^2.
	\label{eq:intscatt2}
\end{equation}
where $\underline{x} = (x,y)$ and $\underline{f} = (f_x,f_y)$. We now define $z_0, z_1, z_2, \ldots, z_N$ to be the boundary profiles between the $N$ layers from the substrate ($z_0 = 0$) to the outer surface. We also denote with $\langle z_0\rangle, \langle z_1\rangle, \langle z_2\rangle, \ldots, \langle z_N\rangle$ the average profile heights. Since the multilayer materials are supposed to be homogeneous, the electron density is constant in each layer and can take on the values ${\cal N}_{\mathrm h}$ or ${\cal N}_{\mathrm l}$, for the high-density or the low-density material, respectively. If the absorption is negligible, also the electric field amplitude is nearly constant in each layer; therefore the integral in $z$ can be solved,
\begin{equation}
	\int _{-\infty}^{z_N} \!\mbox{d}z \, E(z) \,{\cal N}(z) \,e^{- i\alpha z} =\frac{1}{i\alpha}\sum_{j = 0}^N\left({\cal N}_{j+1}E_{j+1} e^{-i\alpha z_{j+1}}- {\cal N}_{j} E_{j} e^{-i\alpha z_{j}}\right).
	\label{eq:intscatt3}
\end{equation} 
Substituting this result into Eq.~\ref{eq:intscatt2}, and re-arranging the terms we obtain
\begin{equation}
	\frac{\mbox{d} I_{\mathrm s}}{\mbox{d} \Omega_{\mathrm s}} =\frac {c}{8\pi} \frac{\mbox{d}\sigma}{\mbox{d}\Omega_{\mathrm s}} \left|\sum_{j =0}^N E_j \,\frac{{\cal N}_{j+1}-{\cal N}_j}{i\alpha}\int_S \mbox{d}^2\underline{x}\: e^{-2 \pi i \underline{f} \cdot \underline{x}} \,e^{- i\alpha z_j} \right|^2,
	\label{eq:intscatt4}
\end{equation}
where ${\cal N}_{N+1} \approx 0$, i.e., the electron density of the ambient. In most cases the contribution of the uppermost interface is negligible, so the difference of electron density at the alternated interfaces can be written as $(-1)^j({\cal N}_{\mathrm h}-{\cal N}_{\mathrm l})$.

Owing to the smooth surface approximation, the exponentials can be approximated as $e^{- i\alpha z_j} \approx e^{- i\alpha \langle z_j\rangle}(1- i \alpha \Delta z_j)$, where $ \Delta z_j = z_j - \langle z_j\rangle$. Substituting and developing the computation \cite{Spiga 2005}, the $0^{th}$ order term yields a delta-like term that represents the ray reflected at $\underline{f} = 0$, i.e., in the specular direction ($\phi_{\mathrm s} =0$, $\theta_{\mathrm s} = \theta_{\mathrm i}$). The subsequent non-zero term describes the scattering at the first order:
\begin{equation}
	\frac{1}{I_0}\frac{\mbox{d} I_{\mathrm s}}{\mbox{d} \Omega_{\mathrm s}} = \frac{({\cal N}_{\mathrm h}-{\cal N}_{\mathrm l})^2}{L_1L_2 \sin\theta_{\mathrm i}} \frac{\mbox{d}\sigma}{\mbox{d}\Omega_{\mathrm s}} \left|\sum_{j =0}^N (-1)^j T_je^{- i\alpha \langle z_j\rangle}\!\int_S \mbox{d}^2\underline{x}\: \Delta z_j \,e^{-2 \pi i \underline{f} \cdot \underline{x}} \right|^2,
	\label{eq:intscatt5}
\end{equation}
where $T_j = E_j/E_0$ and the diagram was normalized to the incident intensity, $I_0 = (c/8\pi)L_1L_2 E_0^2\sin\theta_{\mathrm i}$. The remaining surface integrals are the Fourier transform of the rough profiles of the layer boundaries, then executing the products the expression becomes
\begin{equation}
	\frac{1}{I_0}\frac{\mbox{d} I_{\mathrm s}}{\mbox{d} \Omega_{\mathrm s}} = \frac{(\Delta{\cal N})^2}{\sin\theta_{\mathrm i}} \frac{\mbox{d}\sigma}{\mbox{d}\Omega_{\mathrm s}}\left[ \sum_{j =0}^N T^2_j\,P_j(\underline{f})+ \sum_{j\neq m}(-1)^{j+m}T_jT_m e^{- i\alpha \Delta_{jm}}C_{jm}(\underline{f})\right].
	\label{eq:intscatt6}
\end{equation}
In Eq.~\ref{eq:intscatt6} we have defined $\Delta_{jm} = |\langle z_j\rangle - \langle z_m\rangle |$, $\Delta{\cal N} = {\cal N}_{\mathrm h}-{\cal N}_{\mathrm l}$. $P_j$ and $C_{jm}$ are the 2D Power Spectral Densities and the Crossed Spectral Densities in the stack, as a function of $f_x$ and $f_y$.

If the roughness is isotropic, in grazing incidence the scattering diagram essentially lies in the incidence plane \cite{Church 1986}: so we integrate Eq.~\ref{eq:intscatt6} over $\phi_{\mathrm s}$. Since the scattering power is essentially concentrated near $\phi_{\mathrm s} \approx 0$, the variation of the cross-section with $\phi_{\mathrm s}$ is negligible. From Eqs.~\ref{eq:freqdef1} and~\ref{eq:freqdef2} we have $\cos\theta_{\mathrm s}\,\mbox{d}\phi_{\mathrm s} \approx \lambda \,\mbox{d}f_y$ if $\phi_{\mathrm s} \approx 0$. So the integration operates on the sole Power Density functions, and we have
\begin{equation}
	\frac{1}{I_0}\frac{\mbox{d} I_{\mathrm s}}{\mbox{d} \theta_{\mathrm s}} = \frac{(\Delta{\cal N})^2\lambda}{\sin\theta_{\mathrm i}} \frac{\mbox{d}\sigma}{\mbox{d}\Omega_{\mathrm s}}\left[P_1(f_x) + P_2(f_x)\right],
	\label{eq:intscatt7}
\end{equation}
where the cross section is evaluated at $(\theta_{\mathrm s}+\theta_{\mathrm i})$ and we have set
\begin{eqnarray}
	P_1(f_x) &=& \sum_{j =0}^N T^2_j\,P_j(f_x) \label{eq:intscatt7a}\\
	P_2(f_x) &=& 2 \sum_{j < m}(-1)^{j+m}T_jT_m \cos(\alpha \Delta_{jm})C_{jm}(f_x) \label{eq:intscatt7b}.
\end{eqnarray}

Using now the expression of the optical constant $\delta$ \cite{Stover 1995}, we can express $\Delta {\cal N}$ in terms of the change of $\delta$ at the interfaces, $\Delta\delta$. Accounting for the angular dependence of the electronic cross section, the proportionality factor in Eq.~\ref{eq:intscatt7} becomes
\begin{equation}
	K(\lambda, \theta_{\mathrm s}, \theta_{\mathrm i}) \approx 4\pi^2 \frac{(\Delta \delta)^2}{\lambda^3\sin\theta_{\mathrm i}} \frac{1+\cos^2(\theta_{\mathrm i}+\theta_{\mathrm i})}{2}.
	\label{eq:intscatt8}
\end{equation}
It is easily checked, e.g. via a series development, that 
\begin{equation}
	1+\cos^2(\theta_{\mathrm i}+\theta_{\mathrm s})\stackrel{\theta_{\mathrm s} \approx \theta_{\mathrm i}}{\approx}(1+\cos^22\theta_{\mathrm i})^{1/2}(1+\cos^22\theta_{\mathrm s})^{1/2}.
	\label{eq:approxcos}
\end{equation}
Finally, using Eq.~\ref{eq:approxcos}, and recalling the approximated form of the Fresnel equation \cite{Stover 1995} for the grazing incidence reflectivity $R_{\mathrm F}^{1/2}(\theta) = \Delta\delta/(2\sin^2\theta)$ at the generic angle $\theta$ beyond the critical one, the proportionality factor turns into 
\begin{equation}
	K(\lambda, \theta_{\mathrm s}, \theta_{\mathrm i}) \approx \frac{16\pi^2}{\lambda^3}\sin\theta_{\mathrm i}\sin^2\theta_{\mathrm s}\left[R_{\mathrm F}(\theta_{\mathrm i})R_{\mathrm F}(\theta_{\mathrm s})\right]^{1/2}.
	\label{eq:intscatt9}
\end{equation}
By substituting the expression of Eq.~\ref{eq:intscatt9} into Eq.~\ref{eq:intscatt7} and dropping the subscript '$x$' in the spatial frequency because of the sample isotropy we eventually obtain Eq.~\ref{eq:XRS}.





\bibliographystyle{model1-num-names}
\bibliography{<your-bib-database>}



\end{document}